\documentstyle[12pt]{article}
\begin{document}

\title{Fermi-Bose Transmutation for Stringlike Excitations of
Maxwell-Higgs Systems}

\author{H. Fort and R. Gambini \\
Instituto de F\'{\i}sica, Facultad de Ciencias, Tristan Narvaja 1674 \\
,11200 Montevideo, Uruguay}

\date{\today}
\maketitle

\begin{abstract}

We show that a closed Nielsen-Olesen string in presence of a
point scalar
source exhibits the phenomenon of Fermi-Bose transmutation.
This provides
physical support to previous claims about transmutation between
bosonic and fermionic one-dimensional structures in (3+1) dimensions.
In order to render the computations mathematically rigorous
we have resorted to an Euclidean lattice regularization.

\end{abstract}


\newpage

        Two different approaches have been followed in order to
prove the (2+1) Fermi-Bose transmutation. First,
Wilczek \cite{w} showed that
the bound state of a quantum point charge and
a classical vortex can have any (fractional) angular momentum.
Secondly, Polyakov \cite{p} proved the fermion-boson
transmutation by considering properties of
an scalar field interacting
with a Chern-Simons term. When the partition function of the scalar
particle represented by a Wilson loop  is computed, the Chern-Simons
action leads to a Gauss linking number.
It turns out that the effect of this linking number contribution
is to turn the bosonic propagator into a spinorial propagator.
In both cases the underlying physical phenomenon
which produces the transmutation of bosons into fermions
is the Ahronov-Bhom \cite{ab} effect.In particular,
the statistics of the
composite system depends on the value of the
product: charge $\times$ magnetic flux.

The two mentioned approaches have been extended to
string-like objects in (3+1) dimensions. It
was shown \cite{gs} that a
composite formed from a Nambu charged closed
string interacting with a
Kalb-Ramond point vortex in a multiply connected three
dimensional space
\footnote{ A torus was excised from the space manifold in
order to
provide stability for the closed strings and the point
source was
located in the forbidden region. A similar construction
was used by
Bowick et al \cite{bghhs}.} can have
fractional angular momentum. On the
other hand, it was also shown that a naive extension of Polyakov's
construction to the (3+1)-dimensional case leads to transmutation
between bosonic and fermionic one-dimensional structures \cite{fgt}.

However, the considered systems in (3+1)-dimensions where based on
rather {\em ad hoc} models.In fact, Nambu strings in (3+1)-dimensions
are not expected to be fundamental objects but they should rather be
considered as effective excitations of some underlying field theory. In
this letter we address the issue of supplying the physical realization
that was lacking.To do that we shall start by considering a field theory
(the Maxwell-Higgs system) whose classical solitonic excitations behaves
as a Nambu string interacting with a Kalb Ramond vortex. Then we shall
quantize the theory by computing the transition amplitude in terms of
the Feynman Path integral, and by making contact with the Polyakov
construction we shall prove the Fermi-Bose transmutation for the
stringlike excitations of the Maxwell-Higgs system.

The action of the system of ref. \cite{gs} was
\begin{equation}
S = S_{NG} +
\frac{1}{2}e \int_{{\tau}_1}^{{\tau}_2} \int_0^{2\pi} d\sigma
B_{\mu \nu}(x) [\dot{x}^\mu x'^\nu -x'^\mu\dot{x}^\nu], \\
\label{eq:GS}
\end{equation}
\begin{equation}
S_{NG} \propto \int_{{\tau}_1}^{{\tau}_2} d\tau \int_0^{2\pi} d\sigma
[(x'\dot{x})^2-x'^2\dot{x}^2]^{1/2},
\label{eq:NG}
\end{equation}
where $S_{NG}$ is the Nambu-Goto bosonic string free action,
$\dot{x}^\mu=\partial^{\mu}x/\partial \tau$, $x'^{\mu}=
\partial^{\mu}x/\partial \sigma$ and $B_{\mu \nu}$
is a rank-2 potential given by
\begin{equation}
B_{ij}=Q \epsilon_{ijk}x_k/r^3, \;\;\;\;\; i=1,2,3, \; \;j=1,2,3,
\label{eq:Bvortex}
\end{equation}
such that it generates the point vortex of
Kalb-Ramond field at the origin:
\begin{eqnarray}
H_{ijk}= Q \epsilon_{ijk}\delta^3(x), \\
\label{eq:vortex}
\! \! \! \! \! \! H_{0ij}=0. \nonumber
\end{eqnarray}
The bosonic action (\ref{eq:NG})
was regarded as an effective
action for solitonic solutions of some underlying theory.
In order to provide such a field theory
our starting point is the well known fact
that the $D=3+1$ Maxwell-Higgs theory,
the relativistic generalization of Ginzburg-Landau
theory of superconductivity,
allows for vortex-line
solutions \cite{no}, namely, the Nielsen-Olesen strings.
Those strings, which
are the analogous to the Abrikosov vortex lines in a
type II superconductor, can be identified
with the Nambu string in the strong coupling limit. Furthermore,
the Maxwell-Higgs classical action:
\begin{eqnarray}
S=\int \frac{1}{2} F\wedge\;^*F &+& \frac{1}{2} (d+ieA)\phi \wedge \;^*(
d-ieA){\bar \phi} - V(\phi) \nonumber\\
\label{eq:MH}
V(\phi)&=& \lambda (|\phi^2| - 1 )^2
\end{eqnarray}
(the $*$ denotes the Hodge dual)
reduces in the strong coupling limit to the Nambu-Goto action
which govern the dynamics of the vortices \cite{f}.

        On the other hand,
writing the scalar field as $\phi = \rho e^{i\varphi }$
we can express the action $S$ as
\begin{eqnarray}
S=\int \frac{1}{2} F\wedge\; ^*F + \frac{1}{2} (d\varphi + eA)\wedge\; ^*
(d\varphi + eA)+ \nonumber \\
+\frac{1}{2}d\rho\wedge\;^*(d\rho) -V(\rho).
\label{eq:S}
\end{eqnarray}
The equation of motion for $A$ and $\varphi$ which follows from this
action,
\begin{equation}
d\;^*F=\;^*(d\varphi+eA)e,
\end{equation}
can be solved by introducing a
two-form field $B$ whose field strength is the three-form field $H$:
\begin{equation}
d\varphi + eA \equiv\;^*(dB) \equiv\;^*H,
\label{eq:B}
\end{equation}
provided that
\begin{equation}
d\,^*H = e F \label{eq:HF}
\end{equation}
Then, using eq.(\ref{eq:HF}),
we can rewrite eq.(\ref{eq:S}) as
\begin{eqnarray}
S \equiv =\int \frac{1}{2} F\wedge\;^*F + \frac{1}{2} H\wedge\;^*H
+e B\wedge F+\nonumber\\
+\frac{1}{2}d\rho\wedge\;^*(d\rho) - V(\rho).
\label{eq:Sd}
\end{eqnarray}
In other words, there is a duality transformation connecting the
scalar field $\varphi$ with a Kalb-Ramond $B$ field.
So, in the strong coupling limit we have three equivalent
classical actions:
(\ref{eq:NG}), (\ref{eq:MH}) and (\ref{eq:Sd}).

By adding to the action (\ref{eq:Sd})
a coupling term with an $external$ 2-form potential $B^{ext}$
given by (\ref{eq:Bvortex}),
\begin{equation}
\delta S = e B^{ext}\wedge F,
\label{eq:deltaS}
\end{equation}
we get an action equivalent to (\ref{eq:GS}).
If we now perform the duality transformation which transforms the
2-form field $B$  into the scalar $\varphi$, this term
becomes
\begin{equation}
e\int j \wedge A \equiv e \oint_C  Q A,
\label{eq:J}
\end{equation}
where $j_\mu(x) =  Q \oint_C \delta(x-y) dy_\mu$ is a
1-form dual to the external 3-form vortex external source.
Notice that we are generalizing the action  (\ref{eq:GS}) by including an
arbitrary vortex, instead of a static vortex.

Thus,
we conclude that the
system consisting of a Maxwell field interacting with a
a dynamical scalar field plus an external source --the
term (\ref{eq:J})--describes (at least in the
strong coupling limit) the composite (\ref{eq:GS}), and will probably
present statistical transmutation at the quantum level.

            In order to consider the quantum treatement
of the preceding vortex excitations we will resort to
the lattice formulation of the Maxwell-Higgs system.
        There are several examples in the lattice field theory
such that a change of variables in the partition function allows for
a formulation of the theory in terms of the physical excitations.
Banks, Kogut and Myerson \cite{bkm} introduced a general tranformation
which gives rise to a description of any abelian gauge theory
with compact variables
in terms of variables on the dual lattice
associated with the topological excitations \cite{bwpp}.
This was a generalization of the
technique used by Jose, Kadanoff, Kirkpatrick and Nelson \cite{jkkn}
to derive the Kosterlitz-Thouless \cite{kt}
phase transition in terms of vortices for the $D=2$ $XY$ model.
Given a $D$ dimensional lattice theory
with abelian compact variables on $c_{k-1}$ cells ($k=1$: sites
and spin theory, $k=2$: links and gauge theory, $k>2$ hypergauge theory),
by means of the $Banks-Kogut-Myerson$ transformation,
one arrives to the `$topological$' expression of the partition
function given by
\begin{equation}
Z_T \propto \sum_{
                \begin{array} {c}\,^*\sigma\\
                              \left(  \partial\,^*\sigma = 0 \right)
                \end{array}
                 }\exp
[-2\pi^2\beta \sum_{c_{D-k-1}}\,^*\sigma \hat{\Delta}\,^*\sigma \, ],
\label{eq:topo}
\end{equation}
where $\,^*\sigma$ denotes an integer variable attached to the
$c_{D-k-1}$ cells of the dual lattice. They correspond to the
closed (due to the constraint $\partial\,^*\sigma = 0$)
world "trajectories" of the $D-k-2$ dimensional topological excitations.
In four dimensions, one has world sheets for $k=1$ and world
trajectories for $k=2$.
The $\hat{\Delta}$ represents the propagator operator.

        Let us now come back to the Maxwell-Higgs theory. For simplicity,
we consider the limit $\lambda \rightarrow \infty$ which freezes
the radial degree of freedom of the Higgs field.This is not a strong
restriction, in fact it is known that
the numerical results already  obtained at
$\lambda=1$ are indistinguishable from the frozen case. Thus we get
a compact dynamical scalar variable, i.e. $\varphi \in (-\pi ,\pi]$
interacting with a non-compact gauge field $A_{\mu}(x) \in (-\infty,\infty)$
\footnote{If one considers compact gauge field i.e. an angle
$\theta_{\mu}(x)$ then, in addition to the
Nielsen-Olesen vortices,
there would occur Dirac magnetic monopoles \cite{bkm} associated
to this second compact variable in the
$topological$ expression of the partition function.}.

This model is known to possess two phases, namely, Higgs
and Coulomb \cite{noncompSQED}. The Higgs phase supports
the closed magnetic vortices.

        The partition function for the Villain \cite{v}
form of the lattice action \footnote{ We choose the
Villain form instead of the ordinary Wilson form only
for simplicity, with the Wilson action it is possible
to repeat all that we do here.}
is given by
\begin{eqnarray}
Z = \int (DA ) \int (D\varphi ) \sum_{ n }\nonumber\\
 \exp
[-\sum_{c_1} \frac{\beta}{2} (dA)^2
-\frac{\kappa}{2} (d \varphi -2\pi n - A)^2 ],
\label{eq:Villain}
\end{eqnarray}
where $DA$ ($D\varphi$) denotes the integral over all link $c_1$
(site $c_0$) variables $A$ (\,$\varphi$\,),
$\beta=\frac{1}{e^2}$ is the gauge
coupling constant, $\kappa$ is the Higgs coupling constant and
$n_(c_1)$ are integer variables defined at the lattice links $c_1$
i.e. $k=1$.
For this lattice model the propagator operator
appearing in the $topological$ representation (\ref{eq:topo})
is given by $\hat{\Delta}
= \frac{1}{\Box+m^2}$, where $m^2=\frac{\kappa }{\beta }$
is the mass acquired by the
gauge field due to the Higgs mechanism.

For $D\!=\!4$ the $c_{D-k-1}$ are plaquettes $c_2$ and
the partition function (\ref{eq:Villain})
can be expressed as a sum over closed surfaces on the dual lattice
\cite{pwz} which are the world sheets of closed
string-like objects.Notice that $k=1$ and therefore $D-k-2=1$).
These closed strings are obtained by intersecting the
closed world sheets with a plane $t= constant$ and can be interpreted
as the lattice version of the classical (continuum) Nielsen-Olesen
magnetic vortices.

        The partition function with an external current loop
of the kind of (\ref{eq:J}), generated by
a charge $Qe$, can be written in the topological representation as
\begin{eqnarray}
Z_T[j_C] \hspace{ 1 cm} \propto \hspace{2 cm} \nonumber\\
\sum_{
                \begin{array} {c} \,^*\sigma(*c_{D-k-1})\\
                              \left(  \partial \,^*\sigma = 0 \right)
                \end{array}
                 }\exp
[-2\pi^2\kappa \sum_{*c_{D-k-1}} \,^*\sigma \frac{1}{\Box+m^{2}}
\,^*\sigma  \nonumber \\
- \frac{Q^2e^2}{2} \sum_{c_k} j_C \frac{1}{\Box+m^{2}} j_C \nonumber \\
-2\pi i Qe\sum_{c_k}  j_C \frac{1}{\Box+m^2} \partial \sigma \nonumber \\
+2\pi i Qe\sum_{c_k} j_C \frac{1}{\Box} \partial \sigma]. \label{eq:W}
\end{eqnarray}

The first three terms in the exponent describe short range (Yukawa)
interactions. The last long range term
is a four-dimensional
analogue of the Gauss linking number for loops in three dimensions
i.e. the linking number $\ell k(\sigma,j_C)$ of world sheets of the strings
and the current $j_C$ which appear in the Wilson loop.

        If we consider the strong coupling limit $\kappa /\beta=m^2
\rightarrow \infty$
we get
\begin{eqnarray}
Z_T[j_C] \hspace{1 cm}\propto \hspace{2 cm}\nonumber\\
\sum_{
                \begin{array} {c} \,^*\sigma(*c_{D-k-1})\\
                              \left(  \partial \,^*\sigma = 0 \right)
                \end{array}
                 }\exp[-2\pi^2\beta \sum_{*c_{D-k-1}} \,^*\sigma^2 \;]
\exp [ 2\pi iQe \sum_{c_k} \ell k(\sigma,j_C) ]
\label{eq:Wsc}
\end{eqnarray}

        This is exactly the form of the Wilson loop average obtained in Ref
        \cite{fgt} from the 3+1 generalization of the Polyakov \cite{p}
        construction that includes a 4-dimensional Chern-Simons interaction.

\begin{equation}
\int  B \wedge F,
\end{equation}
 The continuous space-time form of this average is

\begin{equation}
Z= \sum_{\sigma} e^{-TS(\sigma)}\exp
[ -\frac{i}{4\pi^2} \oint_C
dx\int_{\sigma_C}d\sigma(x')^*
( d\frac{1}{|x-x'|^2})\,
].
\label{eq:cont}
\end{equation}

        As in Ref.\cite{p}, the integral in the exponential is a
topological number: the linking number $\ell k(\sigma,C)$ which
measures the number of times the closed path $C$ intersects the sheet
$\sigma$ in four dimensions.There is however an important difference
between this expression obtained from the 3+1 extension of the
Polyakov's construction and the lattice expression (\ref{eq:Wsc}).
As in the 2+1 case, the Polyakov's construction leads to a
singular expression, since C must tend to be the
border of $\sigma$, and some
regularization process is needed, while in the lattice version
the previous singularity has disappeared.
        From here, one can immediately prove the transmutation by
following the same steps that in \cite{fgt}. The main idea is that the
action appearing in (\ref{eq:cont}) leads after a Dirac quantization to
a set of variables that behave as Pauli matrices and reproduce the
propagator of the fermionic string.

Notice that in order to recover the specific coefficient appearing in
(\ref{eq:cont}) one has to suitable tune the value of Qe.
This is in agreement with \cite{gs}, where the Aharonov-Bohm effect
produces the transmutation to a fermionic string only for certain values
of $Qe$.

One could have followed a similar approach in the 2+1 case.
Starting from the charge vortex pair, one could have studied the
transition amplitude by computing the Feynman path integral of a charged
particle in the presence of a magnetic vortex. This computation would
have led to the ordinary Gauss linking number and to the action already
considered by Polyakov in ref[\cite{p}].

So, we have proved that the stringlike excitations of the Maxwell-Higgs
system
in presence of an external charge undergo Fermi-Bose transmutation. It
still remains to be understood if this kind of mechanism have some
physical relevance, in particular if it may affect the behavior of the
high temperature superconductors.

\end{document}